\pdfoutput=1
\documentclass[conference, a4paper, 9pt]{IEEEtran}
\usepackage[english]{babel}
\usepackage[T1]{url}
\usepackage[final]{microtype}
\usepackage{amsmath}
\usepackage[romanConstants, boldVectors]{scstuff}
\usepackage{units}
\usepackage{acronym}
\usepackage{cite}
\usepackage[caption=false, subrefformat=parens]{subfig}
\usepackage[pdftex]{graphicx}
\usepackage{booktabs}
\usepackage[inline]{enumitem}
\usepackage[pdftex, unicode, bookmarks=false, bookmarksnumbered=true,
  pdfpagemode={UseOutlines}, plainpages=false, pdfpagelabels=true,
  colorlinks=true, linkcolor={black}, citecolor={black}, 
  urlcolor={black}]{hyperref}
\usepackage{ucs}
\usepackage[utf8x]{inputenc}
\usepackage{autofe}
\usepackage{mdwtab}
\usepackage[textsize=footnotesize, textwidth=1cm]{todonotes}
\makeatletter
\let\thanks\@IEEESAVECMDthanks
\makeatother

\PrerenderUnicode{©}

\renewcommand\normalsize{\fontsize{9}{11}\selectfont}

\graphicspath{{./Figures/}}

\makeatletter
\newcommand\isacroused[3]{%
  \expandafter\ifx\csname ac@#1\endcsname\AC@used
  #2\else #3\fi}
\makeatother
\newacro{EMC}{Electro-Magnetic Compatibility}
\newacro{PWM}{Pulse Width Modulation}
\newacro{CE-SS}{Constant Envelope, Spread Spectrum}
\newacro{FM}{Frequency Modulation}
\newacro{PAM}{Pulse Amplitude Modulation}
\newacro{PSD}{Power Spectral Density}
\newacro{ESD}{Energy Spectral Density}
\newacro{PDF}{Probability Density Function}
\newacro{SLSQP}{Sequential Least Squares Quadratic Programming}
\newacro{BP}{Band-Pass}
\newacro{LP}{Low-Pass}

\newcommand{\Df}{\Delta f}
\newcommand{\spectrum}[1]{\ensuremath{\Phi_{#1#1}}}
\newcommand{\barspectrum}[1]{\ensuremath{\bar\Phi_{#1#1}}}

\title{Achievement of Preassigned Spectra in the Synthesis of\\
  Band-Pass Constant-Envelope Signals by\\
  Rapidly Hopping through Discrete Frequencies}%
\author{%
  \IEEEauthorblockN{Sergio Callegari}
  \IEEEauthorblockA{ARCES/DEI, University of Bologna, Italy\\
    \mailto{sergio.callegari@unibo.it}}%
  \thanks{This is a post-print version of a paper published in the Proceedings
    of the 2014 IEEE International Symposium on Circuits and Systems (ISCAS
    2014).  Available via DOI \doi{10.1109/ISCAS.2014.6865749}.
    Cite as:\protect\\[1ex]
    Callegari, S., ``Achievement of preassigned spectra in the synthesis of
    band-pass constant-envelope signals by rapidly hopping through discrete
    frequencies,'' 2014 IEEE International Symposium on Circuits and Systems
    (ISCAS), pp.~2776--2779, Jun.\@ 2014.
    \protect\\[1ex]
    Copyright © 2014 IEEE. Personal use of this material is permitted. However,
    permission to use this material for any other purposes must be obtained
    from the IEEE by sending a request to \mailto{pubs-permissions@ieee.org}.
    \protect\\[-2ex]} }

\initializeplaintitle
\initializeplainauthor[%
\def\IEEEauthorblockN#1{#1}%
\def\IEEEauthorblockA#1{}]
\hypersetup{pdftitle=\plaintitle, 
  pdfauthor=\plainauthor,
  pdfsubject={Synthesis of random-FM signals using discrete frequencies},
  pdfkeywords={Random-FM modulation, spread spectrum, chaos, Optimization}}

\begin{document}
\maketitle

\begin{abstract}
  Spread-spectrum signals are increasingly adopted in fields including
  communications, testing of electronic systems, \ac{EMC} enhancement,
  ultrasonic non-destructive testing. This paper considers the synthesis of
  constant-envelope band-pass waveforms with preassigned spectra via an
  \acs{FM} technique using only a limited number of frequencies. In particular,
  an optimization-based approach for the selection of appropriate modulation
  parameters and statistical features of the modulating waveform is
  proposed. By example, it is shown that the design problem generally admits
  multiple local optima, but can still be managed with relative ease since the
  local optima can typically be scanned by changing the initial setting of a
  single parameter.
\end{abstract}
\acresetall

\section{Introduction}
In recent years, signal processing techniques exploiting spread-spectrum
signals have received increasing attention, pushed by the development of novel
communication schemes \cite{Scholtz:TOC-1982}. Yet, other significant
applications exist, including: the testing of analog circuits or communication
channels \cite{pan:TCAD-16-10, negreiros:DATE-2004, Callegari:TCAS1-57-5}; the
enhancement of \ac{EMC} in clocked systems or in \ac{PWM} \cite{Hardin:EC94,
  Balestra:TIEICEE-E87C-1}; the reduction of noise in auto-zero amplifiers
\cite{Tang:ISCAS01}; non-destructive ultrasound testing with coded excitations
and the pulse-compression approach \cite{Gan:Ultrasonics-39-3}.

In this context, the design of sources delivering \ac{CE-SS} signals is a
particularly interesting problem. Constant envelope is relevant where power
delivery is a key issue, letting amplifiers work close to their maximum
specifications, and so making an efficient use of hardware and energy. For
instance, \ac{CE-SS} signals are directly adopted in communication schemes such
as FM-DCSK \cite{Kennedy:CET-2000} or in ultrasound testing
\cite{Callegari:IUS-2012}. Furthermore, \ac{CE-SS} signals can often be
post-processed into spread-spectrum clocks \cite{Pareschi:ICBC-7-2009} and
PWM-like waves for DC–DC converters, motor drives and audio drives
\cite{Balestra:TIEICEE-E87C-1}.

The applicability of \ac{CE-SS} sources depends on the ease of implementation
in integrated form and in the possibility of tuning them for different
requirements. Specifically, the ability to deliver a pre-assigned output
spectrum is significant in tasks such as \ac{EMC} enhancement (where maximally
flat spectra are sought), ultrasound (where spectra matching the probe response
can be useful \cite{Callegari:IUS-2012}), or analog testing. Clearly, spectrum
shaping cannot be practiced by linear filtering as this would hinder the
constant envelope property and must thus be inherent in the signal generation
process.

\begin{figure}
  \begin{tightcenter}
    \includegraphics[scale=0.75]{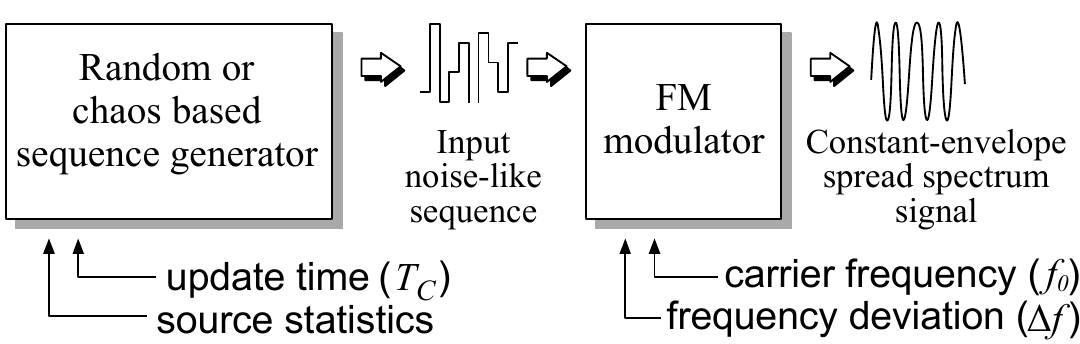}
  \end{tightcenter}
  \caption{Architecture of \ac{FM} based \ac{CE-SS} signal generator.}
  \label{fig:arch}
\end{figure}

A convenient way to generate \ac{CE-SS} \ac{BP} signals consists in feeding a
random or chaotic \ac{PAM} sequence into a \ac{FM} block, as in
Fig.~\ref{fig:arch}. This architecture is easily implementable and mathematical
tools exist for the \emph{analysis} of the achieved spectral features
\cite{callegari:TCAS1-50-1}, both for the random and the chaotic case (even if
the latter may introduce specific features \cite{callegari:TCAS1-50-1,
  Callegari:EL-38-12}). In this paper, the matter of reversing the analysis
tools into \emph{design methods} is considered. This has so far been tackled
only for specific combinations of \ac{FM} parameters and target spectra. For
instance, tools exist for modulations where one \emph{slowly} hops through
frequencies picked through a continuous valued \ac{PAM} sequence
\cite{Callegari:IJCTA-30-5} or for flat goal spectra
\cite{Pareschi:ICBC-7-2009}.  Here, the target are \emph{fast} modulations
where one can only hop through a limited number of frequencies and an arbitrary
goal \ac{PSD} can be specified. The problem is interesting for two main
reasons:
\begin{enumerate*}[label=(\roman*)]%
\item in many practical applications spectral features need to be evaluated on
  relatively short time spans where a slow hopping may result in a too limited
  number of tones being excited;
\item relying on a reduced set of tones may simplify the \ac{PAM} sequence
  generator and the \ac{FM} block.
\end{enumerate*}

The key of the current proposal consists in formulating the choice of the
\ac{FM} parameters and the modulating sequence statistics in a form manageable
by a nonlinear optimizer \cite{kraft:TMS-20-3}. It can be experimentally
observed that the design problem has multiple local optima. Yet, typically,
these can be easily scanned by changing a single \ac{FM} parameter in the
initialization vector for the optimizer. Interestingly, for fast modulations,
the optimization may often end up switching off some tones completely, so that
the number of used frequencies can eventually be even lower than initially
devised.

\section{Background}
\label{sec:background}
In Fig.~\ref{fig:arch}, the \ac{FM} block is continuous-phase. The signal fed
into it shall be indicated as $x(t)$. Being a \ac{PAM} signal, it can be
expressed via a sequence $x_k$, so that $x(t)=x_k$ for $t\in[k T_C, (k+1) T_c[$
where $T_c$ is the \ac{PAM} update period. Values $x_k$ can be assumed to lie
in $[-1,1]$.  The \ac{FM} control parameters are the center frequency $f_0$ and
the deviation $\Df$ (so that when $x(t)$ spans $[-1,1]$, frequency spans
$[f_0-\Df,f_0+\Df]$). With no loss of generality, the constant envelop \ac{FM}
signal $s(t)$ can also be assumed to lie in $[-1,1]$, so that
\begin{equation}
  s(t)=\cos\left(2\pi f_0 t + 2\pi \Df \int_{-\infty}^{t} x(t) dt \right) ~. 
\end{equation}
With this, its cycle-averaged power is fixed at $\nicefrac{1}{2}$.  As long as
$x_k$ is continuously distributed and aperiodic, one can expect the output
\ac{PSD} to be continuously distributed in a \ac{BP} frequency range.
Particularly, if values $x_k$ are independent from each other, the output
spectrum has been shown \cite{callegari:TCAS1-50-1, Callegari:IJCTA-30-5} to be
given by
\begin{multline}
  \label{eq:random-fm}
  \spectrum{s}(f)=
  \int_{-1}^1 K_1(x,f-f_0)\rho(x)\,dx +\\
  \real\left(
    \frac{%
      \left(
        \int_{-1}^1 K_2(x,f-f_0)\rho(x)\, dx
      \right)^2}
    {1-\int_{-1}^1 K_3(x,f-f_0)\rho(x)\, dx}
  \right) ~.
\end{multline}                                                                 
In this expression, $\rho(x)$ is the \ac{PDF} of the modulating waveform and
the three kernels are defined as
\begin{equation}
  \begin{split}
    \label{eq:kernels}
    K_1(x,f)&=\frac{1}{2} \frac{m}{\Df} \sinc^2\left(\pi \frac{m}{\Df}(f
      -\Df\, x)\right)\\
    K_2(x,f)&=\ii \frac{\ee^{-\ii 2 \pi \frac{m}{\Df} (f-\Df\,x)}-1}{2\pi
      \sqrt{\frac{m}{\Df}} (f-\Df \,x)}\\
    K_3(x,f)&=\ee^{-\ii 2 \pi \frac{m}{\Df} (f-\Df x)}
  \end{split}
\end{equation}
where a modulation index $m=\Df T_c$ is introduced for convenience. A large $m$
results in a \emph{slow} modulation, namely, the speed of frequency changes
$\nicefrac{1}{Tc}$ is low compared to $\Df$. For very large $m$ values,
$\spectrum{s}(f)$ tends to take the same shape as $\rho(x)$, while for lower
$m$ and faster modulations the \ac{PSD} ends up taking a nonlinearly
\emph{smoothed} version of the shape of $\rho(x)$.

At first sight, slow modulations may look appealing since they ease the
spectrum shaping of $s(t)$. Indeed, they allow $f_0$, $\Df$ and the \ac{PDF}
$\rho(x)$ to be trivially chosen to obtain any desired \ac{BP}
\ac{PSD}. However, Eqn.~\eqref{eq:random-fm} holds for infinitely long signals
while in most applications behaviors are observed through finite time
spans. For a short signal chunk, the \ac{ESD} can differ from $\spectrum{s}(f)$
and this is particularly true for slow modulations, where the frequency hopping
mechanism can be easily perceived by observing $s(t)$. Additionally, in a
finite time $T_o$ the number of frequency changes $n_c=\nicefrac{T_o}{T_c}$ can
be low, leading to an \ac{ESD} with just a few peaks. On the contrary, in fast
modulations a \emph{frequency merging} occurs and $s(t)$ can be seen as
simultaneously stimulating a whole range of frequencies at any given
time. Furthermore, since $f_0$ and $\Df$ are generally pre-assigned, lowering
$m$ means lowering $T_C$ and thus enlarging $n_c$.

For these reasons, fast modulations should actually be preferred. As a further
advantage, they let the frequency merging phenomenon be exploited to achieve
continuous \ac{PSD} using only a few discrete values in $x_k$ (namely only a
few tones). However, deployment is difficult. Recently, some techniques have
been developed to design $f_0$, $\Df$ and $\rho(x)$ at relatively low $m$
values \cite{Callegari:IJCTA-30-5}, yet without reaching situations where $x_k$
could be discrete valued. Alternatively, $m$ has been optimized to work with
binary balanced random or chaotic $x_k$, but only to obtain \emph{flat}
\acp{PSD} \cite{Callegari:EL-38-12}.

\section{Selection of an optimal random modulating sequence}
\label{sec:optimal}
Here, the problem of using rather fast (actually optimally fast) modulations to
deliver pre-assigned \acp{PSD} out of a limited set of tones is considered. To
start, note that to approximate a pre-assigned \ac{BP} \ac{PSD}
$\barspectrum{s}(f)$, one wants $f_0$, $\Df$, $m$ and $\rho(x)$ to be chosen so
that
\begin{equation}
  \nu = \int_{-\infty}^{\infty} \abs{\barspectrum{s}(f)-\spectrum{s}(f)} df
  \label{eq:mf}
\end{equation}
is minimized. Intuitively, this involves setting $f_0$ at the center frequency
of $\barspectrum{s}(f)$ and $\Df$ at half of its bandwidth (or slightly less,
considered that in fast modulations some energy necessarily leaks out of the
$[f_0-\Df, f_0+\Df]$ interval). Thus the problem can be reduced to picking the
best $m$ and $\rho(x)$.

If $x_k$ is discrete valued with $N$ levels $L_1, \dots L_N$, one has
\begin{equation}
  \rho(x)=\sum_{i=1}^N P_i\, \delta(x-L_i)
\end{equation}
where $\delta$ is the Dirach delta and $P_i$ is the probability of finding
$x_k$ at $L_i$.  With this, the expression of $\spectrum{s}(t)$ can be
simplified into
\begin{multline}
  \spectrum{s}(f)=
  \sum_{i=1}^N K_1(L_i, f-f_0) P_i + \\
  \real\left( \frac{\left(\sum_{i=1}^N K_1(L_i, f-f_0) P_i\right)^2}{%
      1-\sum_{i=1}^N K_1(L_i, f-f_0) P_i} \right) ~.
  \label{eq:disc-RFM}
\end{multline}
When Eqn.~\eqref{eq:disc-RFM} is plugged into~\eqref{eq:mf}, one gets a merit
factor $\nu(P_1, \dots, P_N, m)$. To ease computation, the integral
in~\eqref{eq:mf} can be limited to some finite interval around $f_0$, as in
$[f_0-f_\gamma, f_0+f_\gamma]$ (for instance with $f_\gamma = 2\Df$). Then
fast, adapting numeric integration algorithms \cite{piessens:quadpack-1983} can
be adopted. With this, the computation of $\nu$ can become fast enough to plug
it into a numeric optimization algorithm, together with the following
constraints
\begin{equation}
  \begin{cases}
    P_i \ge 0 \quad \text{for } i \in \Nset{Z}\cap [1,N]\\
    \textstyle \sum_{i=1}^N P_i = 1\\
    m > 0
  \end{cases} ~.
\end{equation}
The nonlinear nature of the merit factor and the lack of an expression for its
Jacobian, restricts the range of adoptable optimizers.  Furthermore, the merit
factor suggests a non convex nature of the problem and the possible existence
of multiple local minima (indeed, this is the case, as the next Section
illustrates).

In order to study the nature of the problem and the local minima distribution,
this work avoids heuristic optimizers based on randomization to escape local
solutions. Conversely, attention is focused on deterministic techniques, taking
as an input an initial condition vector usable as a selector to explore the
solution space. In all the tests performed to validate the approach, the
\ac{SLSQP} method and associated code \cite{kraft:TMS-20-3} have been adopted,
showing suitability for the problem and good performance. \ac{SLSQP} can manage
both equality and inequality constraints and can estimate the Jacobian of the
cost function autonomously. In case of multiple local minima, the initial
condition determines which one is found. Clearly, a convenient way to study
local minima is to start with a reasonable reference initial condition and then
to apply variations to it.

To build a reference vector of probabilities to be used as initial conditions
one may introduce a sequence $\alpha_i$ with $N+2$ entries, as in
\begin{equation}
  \begin{cases}
    \alpha_0 &= f_0-f_\gamma\\
    \alpha_i &= f_0+L_i \Df \quad \text{for}\ i=1,\dots,N\\
    \alpha_{N+1} &= f_0+f_\gamma
  \end{cases} ~.
\end{equation}
Then, another sequence $\beta_i$ can be obtained as
$\beta_i=\nicefrac{(\alpha_i+\alpha_{i+1})}{2}$, for $i=0,\dots,N$. With this,
one can define
\begin{equation}
  \hat P_i = \int_{\beta_i}^{\beta_{i+1}} \barspectrum{s}(f) df
\end{equation}
and eventually obtain an initial vector of probabilities by normalizing each
$\hat P_i$ over $\sum_{i=0}^{N} \hat P_i$. The rationale for this initial
vector is the following: it always respects the constraints; for $N\to\infty$
it directly leads to the optimal $\rho(x)$ at large $m$ as evident from
analyzing Eqn.~\eqref{eq:disc-RFM} and the kernels in Eqn.~\eqref{eq:kernels};
at moderate $N$ and smaller $m$ it provides spectra typically oscillating
around a smoothed version of the desired one, which both intuitively and
empirically proves to be a reasonable choice.

\section{Simulations, examples and discussion}
\label{sec:examples}
\begin{figure}
  \begin{tightcenter}
    \subfloat[\label{sfig:goal1}]{%
      \includegraphics[scale=0.55]{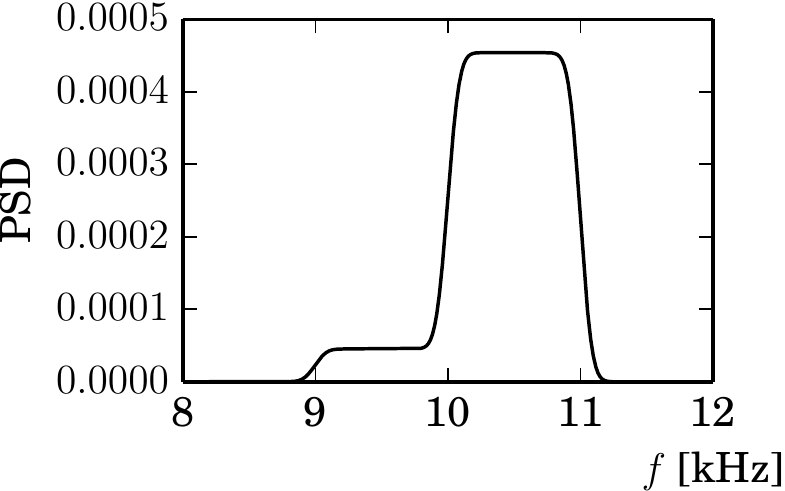}} \hfill
    \subfloat[\label{sfig:initial1}]{%
      \includegraphics[scale=0.55]{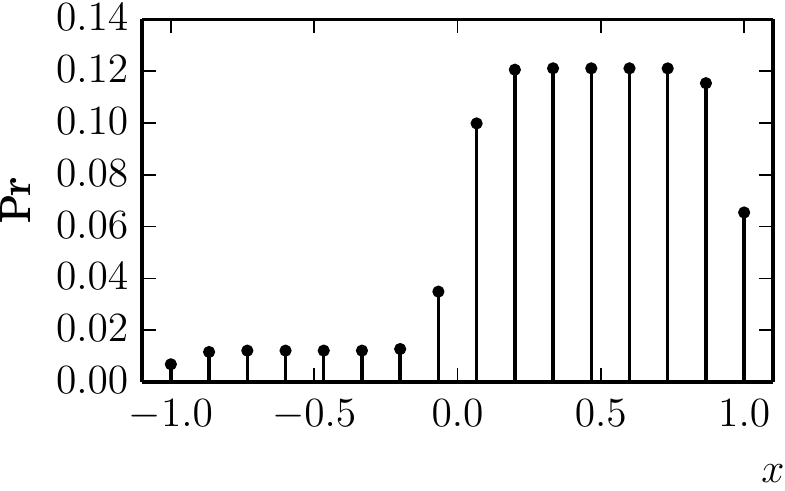}}
  \end{tightcenter}
  \caption{Sample goal \ac{PSD} \protect\subref{sfig:goal1} and corresponding
    reference initial probability vector \protect\subref{sfig:initial1} for
    $N=16$.}
  \label{fig:goal1}
\end{figure}

To discuss the approach, it is worth considering an example.  Assume that the
goal \ac{PSD} is as shown in Fig.~\ref{sfig:goal1}. This is obtained from a
function returning $1$ for $f\in\unit[{[9,10]}]{kHz}$, $10$ for
$f\in\unit[{[10,11]}]{kHz}$ and zero elsewhere. The function is smoothed a
little with a non-causal low-pass filter and then scaled to return
$\nicefrac{1}{2}$ as its integral according to the Parseval theorem.

In the proposed experiments, the discrete levels of the modulating \ac{PAM}
signal are assumed to distribute uniformly in $[-1,1]$. Namely,
$L_i=-1+\nicefrac{2(i-1)}{(N-1)}$ Fig.~\ref{sfig:initial1} shows the reference
initial probability vector for $N=16$.

\begin{figure}
  \begin{tightcenter}
    \subfloat[\label{sfig:prob1.m2.ref}]{%
      \includegraphics[scale=0.55]{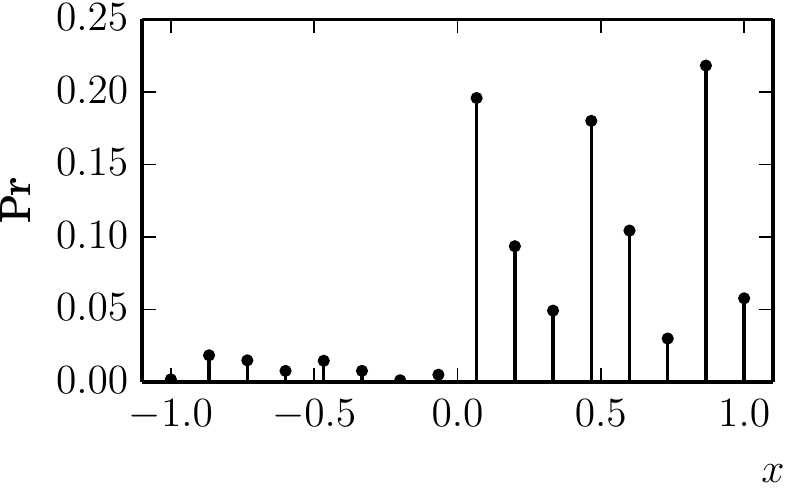}}
    \hfill
    \subfloat[\label{sfig:psd1.m2.ref}]{%
      \includegraphics[scale=0.55]{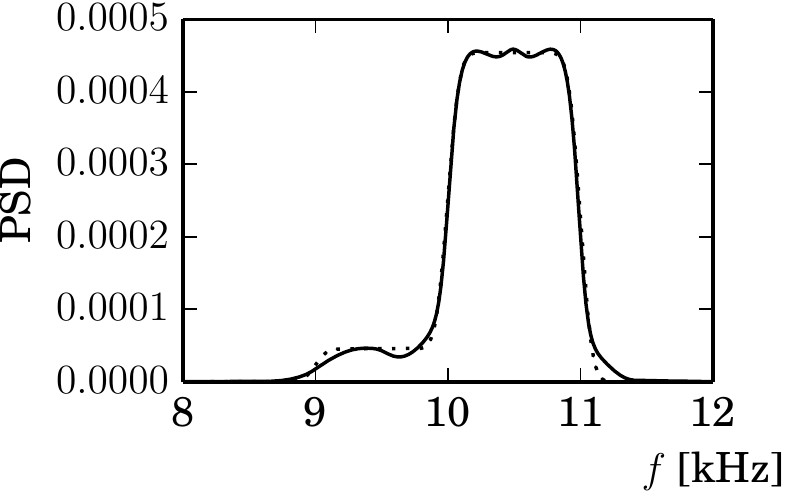}}\\
    \subfloat[\label{sfig:prob1.m2.rnd1}]{%
      \includegraphics[scale=0.55]{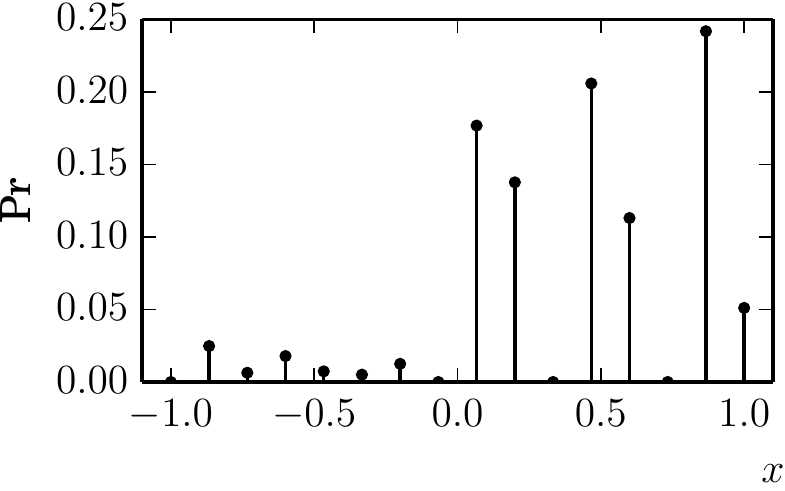}}
    \hfill
    \subfloat[\label{sfig:psd1.m2.rnd1}]{%
      \includegraphics[scale=0.55]{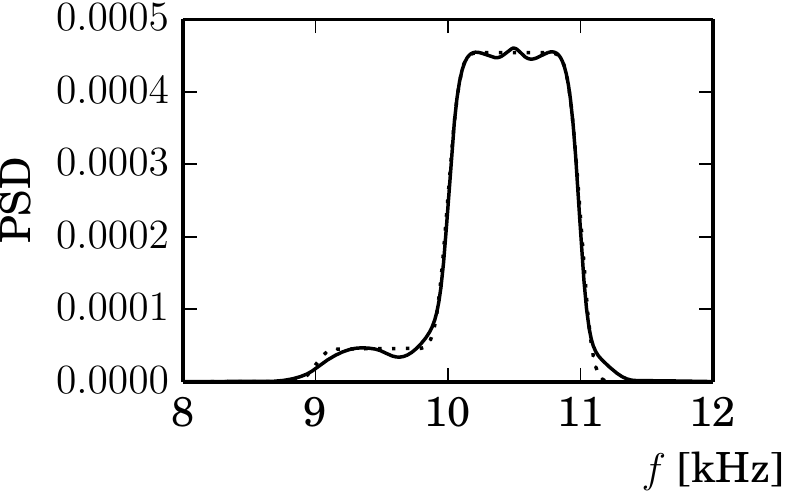}}
  \end{tightcenter}
  \caption{Optimal $P_i$ (left column) and corresponding \ac{PSD} (right) when
    practicing optimization at fixed $m=2$, with $N=16$. In
    \protect\subref{sfig:prob1.m2.ref} and \protect\subref{sfig:psd1.m2.ref}
    the reference initial vector is adopted. The corresponding $\nu$ is
    $0.0182$. In \protect\subref{sfig:prob1.m2.rnd1} and
    \protect\subref{sfig:psd1.m2.rnd1}, a random initial vector is used. $\nu$
    is $0.0171$.}
  \label{fig:fixed-m}
\end{figure}

To begin with, some experiment can be run \emph{fixing} $m$ (preventing the
optimizer from changing it) and merely perturbing the initial probability
vector with respect to the reference one. In this setup, changing the initial
probability vector can in some cases change the optimal solutions being
found. However, all the so found solutions tend to have very little differences
in cost. Furthermore, the achieved $P_i$ values end up being relatively similar
among different solutions, namely, those values that are large in one solution
remain large in another.  Even if experimental tests cannot provide definitive
answers, one can thus conjecture that at fixed $m$, either there are no local
minima (and the observed one are artifacts from finite machine precision) or
they are rather close to each other. As an example, of the few experienced
cases where randomizing the initial condition has resulted in slightly
different solutions is shown in Fig.~\ref{fig:fixed-m}, which refers to $m=2$.

\begin{figure}[t]
  \begin{tightcenter}
    \subfloat[\label{sfig:prob1.ms6.ref}]{%
      \includegraphics[scale=0.55]{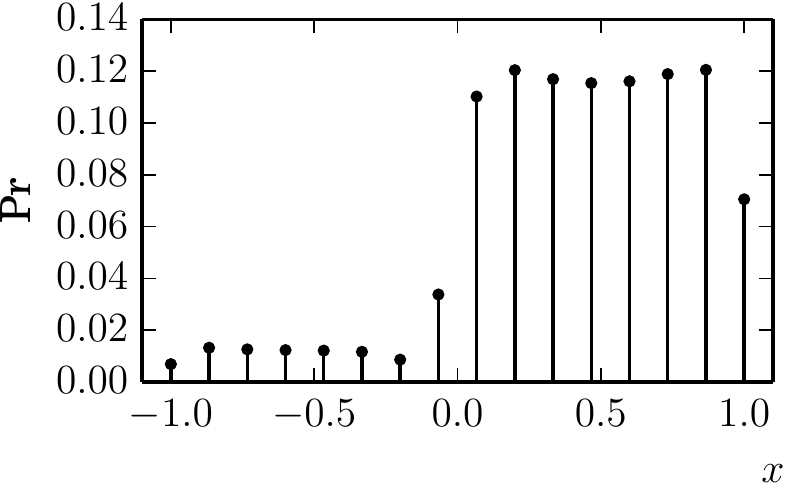}}
    \hfill
    \subfloat[\label{sfig:psd1.ms6.ref}]{%
      \includegraphics[scale=0.55]{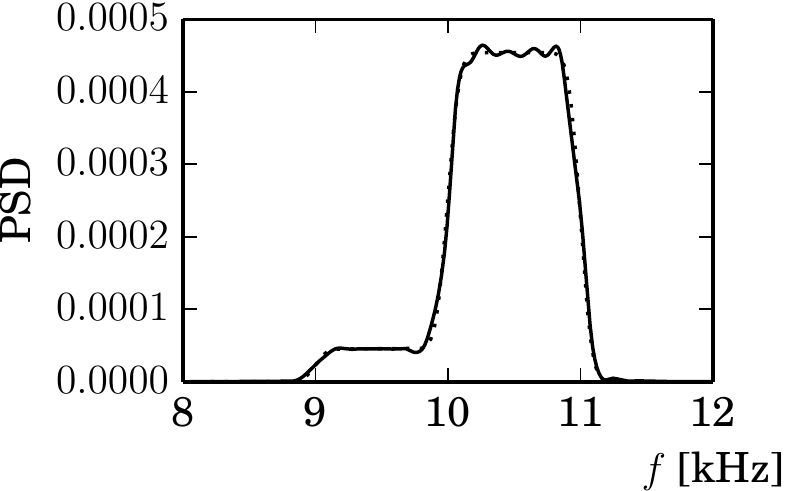}}\\
    \subfloat[\label{sfig:prob1.ms4.ref}]{%
      \includegraphics[scale=0.55]{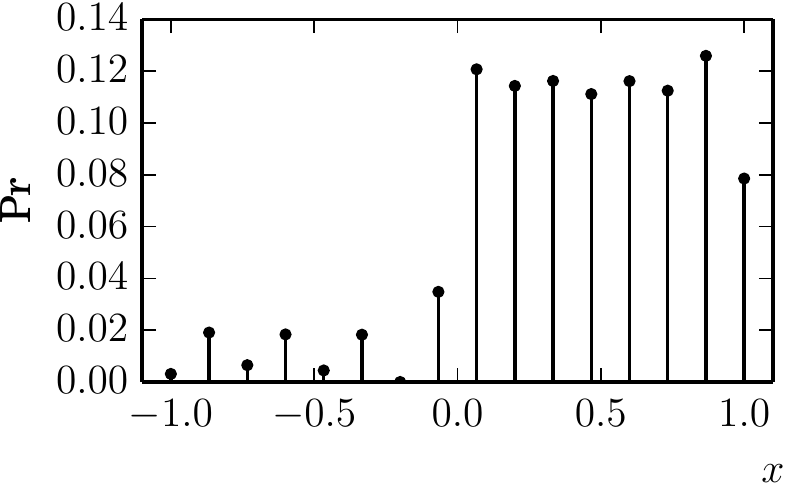}}
    \hfill
    \subfloat[\label{sfig:psd1.ms4.ref}]{%
      \includegraphics[scale=0.55]{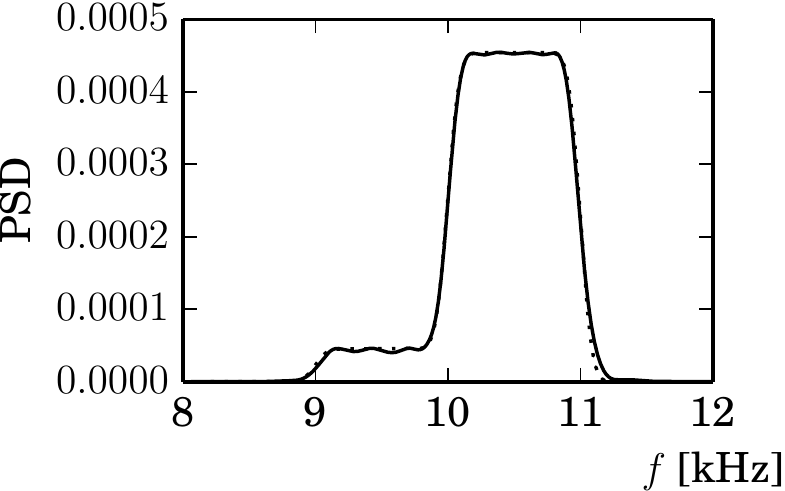}}\\
    \subfloat[\label{sfig:prob1.ms3.ref}]{%
      \includegraphics[scale=0.55]{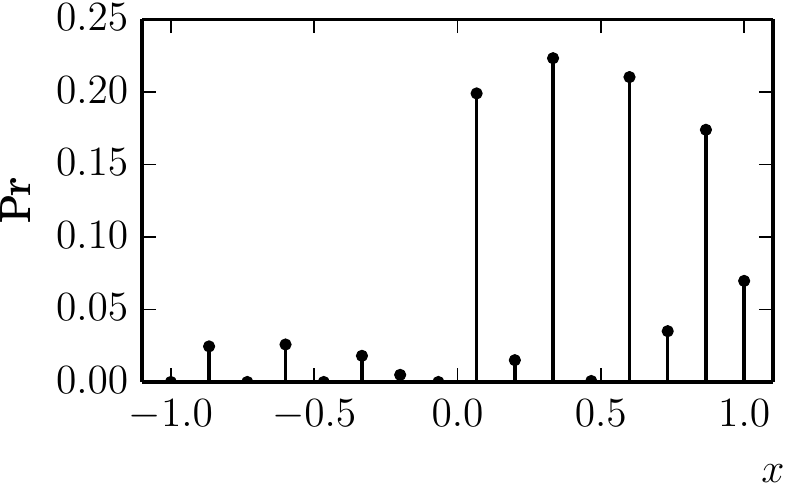}}
    \hfill
    \subfloat[\label{sfig:psd1.ms3.ref}]{%
      \includegraphics[scale=0.55]{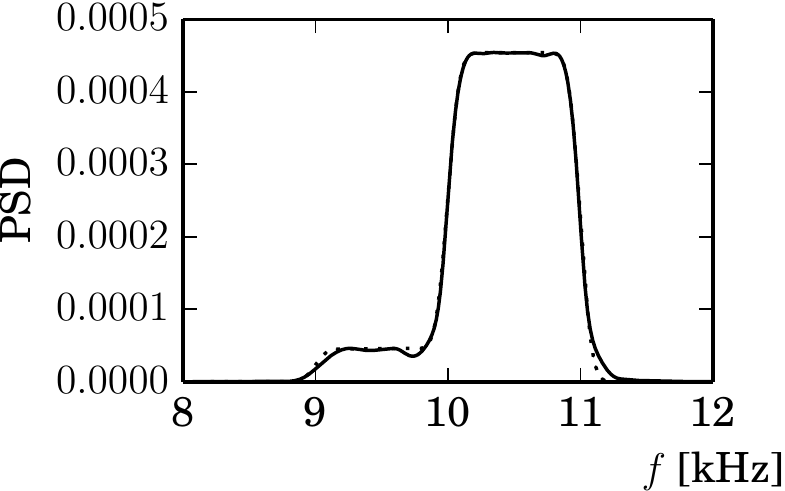}}\\
    \subfloat[\label{sfig:prob1.ms1-5.ref}]{%
      \includegraphics[scale=0.55]{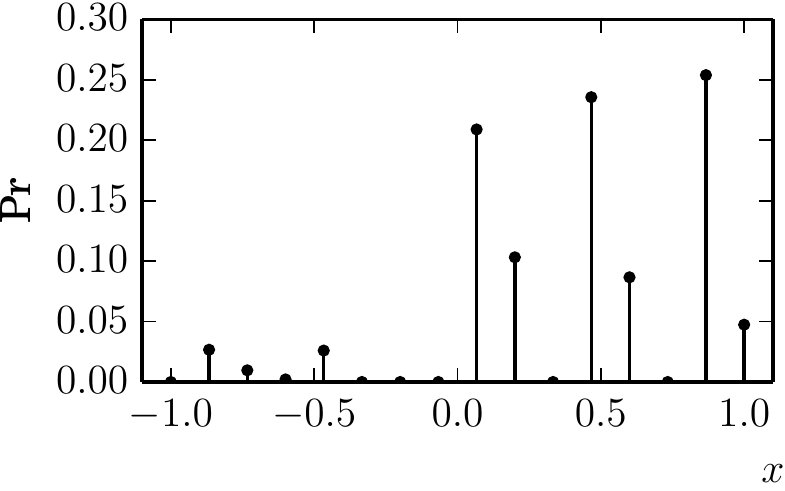}}
    \hfill
    \subfloat[\label{sfig:psd1.ms1-5.ref}]{%
      \includegraphics[scale=0.55]{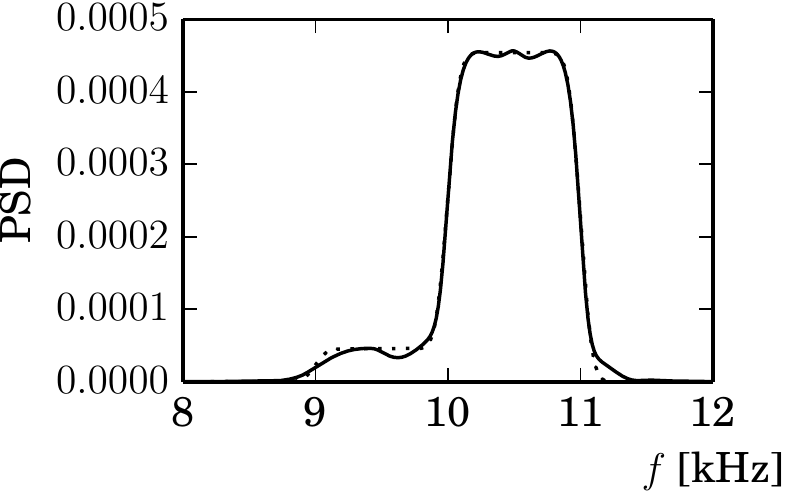}}\\
    \subfloat[\label{sfig:prob1.ms0-2.ref}]{%
      \includegraphics[scale=0.55]{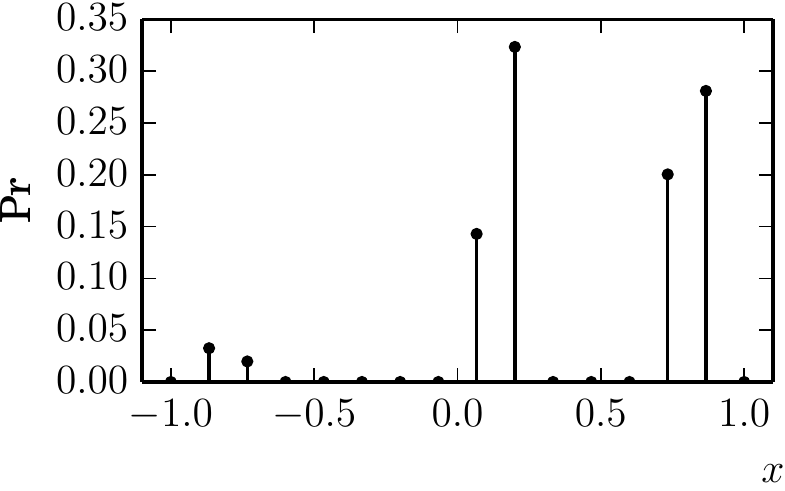}}
    \hfill
    \subfloat[\label{sfig:psd1.ms0-2.ref}]{%
      \includegraphics[scale=0.55]{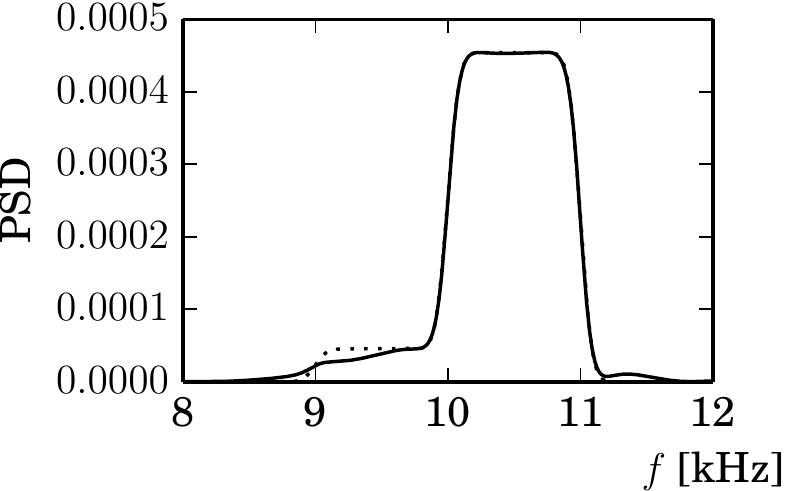}}
    \end{tightcenter}
    \caption{Some local optima. In~\protect\subref{sfig:prob1.ms6.ref}
      and~\protect\subref{sfig:psd1.ms6.ref}, $m=5.72$ and $\nu=0.0146$.
      In~\protect\subref{sfig:prob1.ms4.ref}
      and~\protect\subref{sfig:psd1.ms4.ref}, $m=3.79$ and $\nu=0.0109$.
      In~\protect\subref{sfig:prob1.ms3.ref}
      and~\protect\subref{sfig:psd1.ms3.ref}, $m=2.86$ and $\nu=0.0102$.
      In~\protect\subref{sfig:prob1.ms1-5.ref}
      and~\protect\subref{sfig:psd1.ms1-5.ref}, $m=1.93$ and $\nu=0.0149$.
      In~\protect\subref{sfig:prob1.ms0-2.ref}
      and~\protect\subref{sfig:psd1.ms0-2.ref}, $m=0.99$ and $\nu=0.0157$.}
    \label{fig:opti-m}
\end{figure}

The situation is much more interesting when the optimizer is allowed to choose
$m$ too. In this case, the initial $m$ appears to be a strong selector for the
final solution. Unfortunately, at least starting from the reference probability
vector, it is impossible to partition $\Nset{R}^+$ in simple regions of
convergence. In other words, it is impossible to identify intervals of $m$
values such as the optimizer always converges to an optimal $m$ \emph{inside
  them}. Still, in general, starting at a large $m$ tends to return a large $m$
solution and viceversa. Some examples are shown in Fig.~\ref{fig:opti-m}.

As it can be seen, even with discrete tones one can approximate a given
\ac{PSD} with extremely good accuracy. For instance, this is the case for the
$m=2.86$ solution where the cost is $0.0102$. Even more interestingly, good
approximations can be obtained even at rather small $m$. For instance, it is
possible to follow relatively well the rapid variation of $\barspectrum{s}(f)$
around \unit[10]{kHz} even at $m\approx 1$, as shown in
Fig.~\ref{sfig:psd1.ms0-2.ref}.

An appealing result of the optimization is that local optima corresponding to
low $m$ values have many tones silenced. With this, the \ac{CE-SS} signal can
be eventually generated out of a very little number of frequencies. For
instance, in the sample case, at $m=0.99$, only 6 tones are used, out of the 16
initially available.

\begin{figure}[t]
  \begin{tightcenter}
    \subfloat[\label{sfig:welch4}]{%
      \includegraphics[scale=0.55]{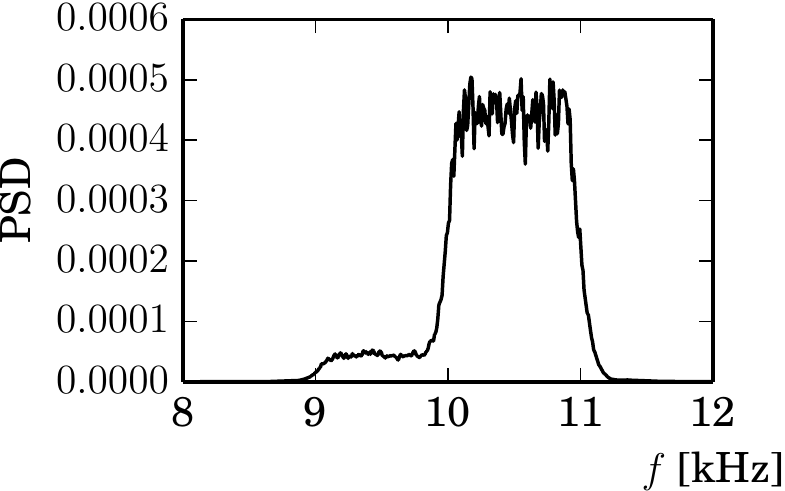}}
    \hfill
    \subfloat[\label{sfig:welch0-2}]{%
      \includegraphics[scale=0.55]{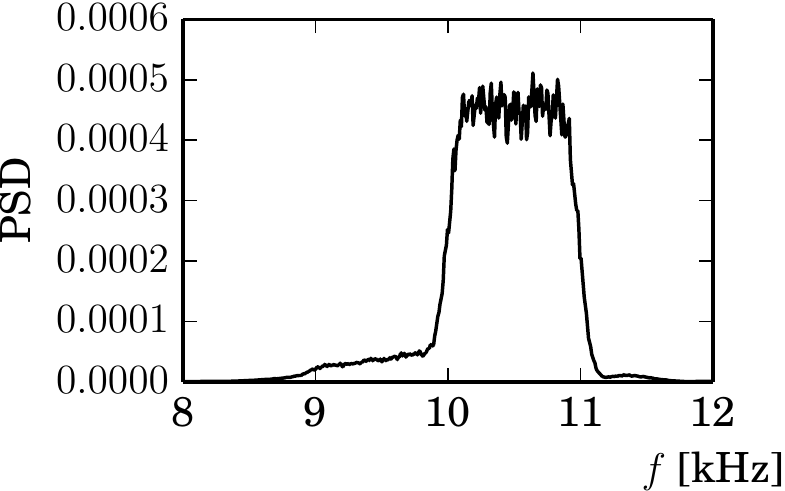}}
  \end{tightcenter}
  \caption{\acp{PSD} obtained from time domain simulations. In
    \protect\subref{sfig:welch4}, \ac{PSD} of a \ac{CE-SS} signal obtained with
    the probability vector in \ref{sfig:prob1.ms4.ref} and with $m=3.79$. In
    \protect\subref{sfig:welch0-2}, the probability vector is that in
    \ref{sfig:prob1.ms0-2.ref} and $m=0.99$.}
  \label{fig:td}
\end{figure}

To show that the approach actually works even when tested for relatively short
signal chunks, Fig.~\ref{fig:td} shows the \acp{PSD} of two \ac{CE-SS} signals
generated with the proposed approach and corresponding to the local minima at
$m=3.79$ and $m=0.99$. Spectral estimation is practiced by the Welch method
operating on signal chunks \unit[16]{s} long. For the estimation the signals
are sampled at $\unit[3.79]{\mu s}$ and the Welch algorithm is tuned to use
windows of 32768 samples. Conformance to the expected \ac{PSD} is almost
perfect in both cases.

\section{Conclusions}
\label{sec:conclusions}
In this work, an optimization based strategy to synthesize \ac{CE-SS} signals
with preassigned spectrum by frequency hopping through limited sets of
available tones has been presented. The approach relies on nonlinear
optimization and has been tested with the \ac{SLSQP} nonlinear optimizer. The
approach can deliver good approximations of the target spectrum, both for the
example case discussed in the paper and for many other that have been tried but
could not be reported. The optimization problem has multiple local minima that
do not represent a significant issue since they can be scanned by selecting
different initial values for the modulation index. Quite interestingly, at fast
modulations the number of tones required for the approximation can turn out to
be much lower than expected.

\section*{Acknowledgment}
\begingroup
Work funded by MIUR PRIN 2009 project ``Diagnostica non distruttiva ad
ultrasuoni tramite sequenze pseudo-ortogonali per imaging e classificazione
automatica di prodotti industriali'' (USUONI).\par
\endgroup

\vfill
\bibliographystyle{SC-IEEEtran}
\bibliography{macros,IEEEabrv,various,sensors,analog,chaos,ultrasound}

\end{document}